\documentstyle[epsf,12pt,aasms4]{article}
\begin{document}
\baselineskip=0.5\baselineskip

\title{The Formation and Structure of a Strongly Magnetized Corona
above Weakly Magnetized Accretion Disks}

\author{Kristen A. Miller and James M. Stone\altaffilmark{1}}
\affil{Department of Astronomy, University of Maryland, College Park, Maryland 20742-2421}

\altaffiltext{1}{also Institute of Astronomy, University of Cambridge,
Madingley Road, Cambridge CB3 0HA, UK}

\begin{abstract}

We use three-dimensional magnetohydrodynamical (MHD) simulations to
study the formation of a corona above an initially weakly magnetized,
isothermal accretion disk.  The simulations are local in the plane of
the disk, but extend up to 5 vertical scaleheights above and below it.
We describe a modification to time-explicit numerical algorithms for
MHD which enables us to evolve such highly stratified disks for many
orbital times.  We find that for initially toroidal fields, or
poloidal fields with a vanishing mean, MHD turbulence driven by the
magnetorotational instability (MRI) produces strong amplification of
weak fields within two scale heights of the disk midplane in a few
orbital times.  Although the primary saturation mechanism of the MRI
is local dissipation, about 25\% of the magnetic energy generated by
the MRI within two scale heights escapes due to buoyancy, producing a
strongly magnetized corona above the disk.  Most of the buoyantly
rising magnetic energy is dissipated between 3 and 5 scale heights,
suggesting the corona will also be hot.  Strong shocks with Mach
numbers $\gtrsim~2$ are continuously produced in the corona in
response to mass motions deeper in the disk.  Only a very weak mass
outflow is produced through the outer boundary at 5 scale heights,
although this is probably a reflection of our use of the local
approximation in the plane of the disk.  On long timescales the
average vertical disk structure consists of a weakly magnetized
($\beta \sim 50$) turbulent core below two scale heights, and a
strongly magnetized ($\beta \lesssim 10^{-1}$) corona which is stable
to the MRI above.  The largescale field structure in both the disk and
the coronal regions is predominately toroidal.  Equating the volume
averaged heating rate to optically thin cooling curves, we estimate
the temperature in the corona will be of order $10^4$~K for
protostellar disks, and $10^8$~K for disks around neutron stars.  The
functional form of the stress with vertical height is best described
as flat within $\pm 2H_z$, but proportional to the density above $\pm
2H_z$.

For initially weak uniform vertical fields, we find the exponential
growth of magnetic field via axisymmetric vertical modes of the MRI
produces strongly buoyant sheets of magnetic energy which break the
disk apart into horizontal channels.  These channels rise several
scale heights vertically before the onset of the Parker instability
distorts the sheets and allows matter to flow back towards the
midplane and reform a disk.  Thereafter the entire disk is
magnetically dominated and not well modeled by the local
approximation.  We suggest this evolution may be relevant to the
dynamical processes which disrupt the inner regions of a disk when it
interacts with a strongly magnetized central object.

\end{abstract}

\keywords{accretion disks -- instabilities -- MHD -- turbulence}

\section{Introduction}

The observed spectrum of thin accretion disks in many systems suggests
that the diffuse upper layers are heated to much higher temperatures
than the midplane, i.e.  that such disks have coronae.  For example,
disks around compact objects (Nandra \& Pounds \markcite{r}1994), CVs
(Yi \& Kenyon \markcite{r} 1997), classical T~Tauri stars (CTTS, Kwan
\markcite{r5}1997), and Herbig AeBe stars (Najita et.  al.
\markcite{r24}1996) all show evidence for the existence of a corona.

One simple explanation for the presence of a corona above a thin disk
is that it is produced by external irradiation from the central object
(or the central, hotter regions of the disk itself).  A variety of
authors have investigated the vertical structure of a thin disk
atmosphere using radiative transfer calculations (Wade \& Hubeny
\markcite{r}1998; Dove et. al. \markcite{r12}1997; Haardt \& Maraschi
\markcite{r}1991) including the effect of irradiation from the central
regions (D'Alessio et.  al.  \markcite{r13}1998; Wood et. al.
\markcite{r}1996).  Such studies often find the diffuse upper layers
of the disk are much hotter than the midplane.  However, it should be
noted that including a sophisticated treatment of the internal
dynamics of the disk into such calculations is a formidable task, thus
important ingredients such as the vertical profile of the heating rate
due to accretion usually must be assumed using, e.g., the
``$\alpha$-disk" prescription (Shakura \& Sunyaev \markcite{r}1973).

A second possible explanation for coronae above accretion disks is
that they are a direct consequence of the internal dynamics of the
disk itself rather than a radiative transfer effect, much in the same
way the solar corona is thought to be heated by dynamical processes
lower in the atmosphere.  This possibility was first suggested as a
mechanism to explain the X-ray emission of Cygnus X-1 (Liang \& Price
\markcite{r1}1977, Galeev et. al.  \markcite{r2}1979) and of Seyfert
AGN (Liang \& Thompson \markcite{r3}1979).  In a detailed
investigation by Stella \& Rosner (\markcite{r4}1984), ``magnetic
buoyancy-driven" convection of strong flux tubes produced by shear in
the disk was identified as the most promising source of the vertical
energy flux needed to heat a corona.  Although it was recognized that
shear amplification of the field would result in stresses on the fluid
that might give rise to angular momentum transport in the disk
(Coroniti 1981; Sakimoto \& Coroniti 1981), and that buoyancy might be
the primary saturation mechanism for the field (Sakimoto \& Coroniti
1989), modeling the coupled amplification, saturation, and angular
momentum transport processes from first principles required
multidimensional MHD models that were beyond the scope of these
pioneering studies.  Nonetheless, these early models were able to fit
observations well enough to suggest that the internal dynamics of the
disk might play an important role in determining the vertical
structure. 

Strong magnetic fields (which have $\beta \le 1$, where $\beta \equiv
8 \pi P/B^{2}$ with $P$ the thermal pressure, and $B$ the magnetic
field strength) in a stratified accretion disk atmosphere are subject
to both the Parker and magnetic interchange instabilities (e.g. Stella
\& Rosner 1984).  As an extension of the earlier work discussed above,
the nonlinear evolution of these instabilities has been the subject of
recent sophisticated numerical MHD modeling (Kamaya et. al.
\markcite{r20}1997; Basu et. al. \markcite{r}1997; Chou
et. al. \markcite{r} 1997; Matsumoto et. al. \markcite{r}1993; Kaisig
et. al. \markcite{r} 1990).  Although such studies have helped clarify
the effect of magnetic buoyancy in determining the vertical structure
of the disk, they do not address the field amplification process
itself, and therefore cannot investigate questions such as how strong
flux tubes are formed in the disk in the first place.  It is clear
that field amplification, buoyancy, and angular momentum transport
processes are all highly coupled, and therefore must be studied in
concert for a complete picture.

In recent years, our understanding of angular momentum transport
processes in accretion disks has progressed rapidly, in part due to
the identification of the magnetorotational instability (MRI) in
weakly magnetized accretion disks, and the realization of the
fundamental role it plays in the process (see the review of Balbus \&
Hawley 1998 for a complete discussion).  Time-dependent MHD
simulations have demonstrated that the MRI saturates as MHD turbulence
in three-dimensions (Hawley, Gammie, \& Balbus 1995a, hereafter HGB1),
and that this turbulence drives a non-helical dynamo which produces
strong amplification of the magnetic field (Hawley, Gammie, \& Balbus
1995b, hereafter HGB2; Brandenburg et al. 1995, hereafter BNST; Stone
et al. 1996, hereafter SHGB; Brandenburg 1998).  Knowledge of the
field amplification process allows us to address the problem of the
formation of a disk corona from first principles.  Tout \& Pringle
(\markcite{r}1992) have presented an interesting analytic argument for
the formation of a heated disk corona via a cyclic process involving
the MRI, the Parker instability, and reconnection.  However, the full
details of such a (nonlinear) cycle require computational studies of
the nonlinear evolution of the MRI in stratified disks.  Current
studies show that while buoyancy does result in a significant vertical
flux of magnetic energy, it is not the dominant saturation mechanism
for the instability: instead local dissipative processes dominate
(BNST, SHGB).  Still, these results hint at the possibility of the
formation of a disk corona via buoyant rise of dynamo amplified
magnetic field in that the scale height of the field was much larger
than that of the gas, leading to a rapid decrease of $\beta$ with
height.

To date numerical simulations of the MRI in stratified disks have been
limited by the relatively small vertical extent that can be included
in the calculations.  This limit is imposed by the enormous range in
dynamical timescales in a stratified disk: the density falls so
rapidly with height in an isothermal disk ($\rho \propto
\exp(-z^{2})$) that the Alfv\'en speed becomes very large in the upper
layers.  For a time-explicit numerical algorithm (in which the
timestep is limited in part by the crossing time of an Alfv\'en wave
across a grid zone), this makes calculations for many orbits of the
disk impracticable with current computer resources.  Thus, the
simulations of BNST and SHGB were limited to two scale heights above
the disk midplane, which was not enough to see the transition from a
weakly magnetized turbulent core to a strongly magnetized corona which
was stable to the MRI.  Moreover, it is not clear what the effect of
the boundary conditions applied at two scale heights may have on the
field strength and geometry in this region.  Ideally, one would like
to apply outflow boundary conditions that allow mass, energy, and
momentum to leave the computational domain, however, as discussed in
SHGB, it is not possible to prescribe a self-consistent outflow
boundary condition for a highly tangled, strong magnetic field without
exerting anomalous stresses on the fluid near the boundary.  Thus, in
most of the simulations described by SHGB, periodic boundary
conditions were applied at the vertical boundaries which, although
they are clearly not a good representation of a real system, are not
significantly less realistic than other viable choices, such as
reflecting walls.  A few simulations presented by SHGB which used
different boundary conditions demonstrated that the vertical profile
of horizontally averaged quantities near the midplane was not changed,
indicating that the vertical structure is relatively independent of
the boundary conditions.  Moreover, the simulations of BNST used
reflecting boundary conditions with a non-zero vertical component to
the magnetic field (so that the turbulence in the disk generates a
time-dependent current on the boundary); the resulting vertical
structure of the disk was not significantly different than that
reported by SHGB.

In this paper, we describe an extension to time-explicit MHD
algorithms which removes the timestep constraint imposed by the
divergence of the Alfv\'en speed in regions of very low density, and
therefore allows numerical simulations over many orbital times of a
disk which spans many scale heights in the vertical extent.  We use
this extension to study the evolution of weakly magnetized, isothermal
stratified disks over many orbital times using a grid which extends 5
scale heights above the disk midplane.  We show that for a variety of
initial field strengths and geometries, buoyantly rising magnetic
field generated by MHD turbulence driven by the MRI near the midplane
of the disk forms a strongly magnetized ($\beta \ll 1$) corona above
two scale heights.  Because it is strongly magnetized, the corona is
stable to the MRI, and moreover the field becomes nearly force-free
there (i.e. $(\nabla \times {\bf B}) \times {\bf B} \approx 0$).
Thus, the field is no longer highly tangled near the upper boundary,
and we are able to apply outflow boundary conditions appropriate to
the system.  These simulations allow us to investigate the formation
and structure of a magnetized corona by calculating from first
principles the coupled field amplification, buoyancy, and angular
momentum transport processes.  We are also able to measure properties
such as the vertical profile of the stress and dissipation rate in the
disk which may be useful in other contexts (such as detailed radiative
transfer calculations of accretion disk atmospheres).  Our models are
still limited by the fact that they are local in the horizontal plane,
which as we discuss may affect processes such as the generation of
magnetocentrifugal winds from the disk surfaces.  However, fully
global models using the methods described here are underway.

The paper is organized as follows: in \S 2 we describe our numerical
methods, in \S 3 we present the evolution of a variety of weak
magnetic field configurations in stratified disks which span many
vertical scale heights, in \S 4 we discuss the implications of our
results for the vertical temperature profiles in CTTS and compact
objects, and in \S 5 we conclude.

\section{Numerical Methods}

We solve the equations of compressible MHD including the effects of
resistivity in the ``shearing box'' approximation (see HGB1 for a
detailed description).  That is, we adopt a frame of reference which
is local, Cartesian, and corotates with the disk at angular frequency
$\Omega$ corresponding to a fiducial radius R$_o$.  In this local
frame, we use spatial coordinates $x=(R-R_o), y=(R\phi-\Omega t)$, and
$z$.  Assuming a Keplerian rotation profile and expanding within a
small volume about the fiducial radius ($\Delta R \ll R_o$), the
equations of motion can be expressed as:
\begin{equation}
 \frac{\partial \rho}{\partial t} + \nabla \cdot 
\rho {\bf v} = 0,
\end{equation}
\begin{equation}
 \frac{\partial {\bf v}}{\partial t} +{\bf v} \cdot \nabla {\bf v}= -
\nabla P/\rho + \frac{1}{\rho c} ({\bf J} \times {\bf B})- 2 \Omega
\times {\bf v} + 3 \Omega^2 x {\bf \hat{x}}-\Omega^2 z {\bf \hat{z}},
\end{equation}
\begin{equation}
 \frac{\partial {\bf B}}{\partial t}=\nabla \times
({\bf v} \times {\bf B}) + \frac{c^2 \eta_0}{4 \pi} \nabla^2 {\bf B},  
\end{equation}
where ${\bf J}$ is the current.  The third and fourth terms on the RHS
in the momentum equation [2] represent Coriolis and tidal forces in
the rotating frame of reference, while gravitational acceleration in
the vertical direction is represented by the last term.  The final
term in the induction equation [3] allows for field dissipation
through an Ohmic resistivity with a constant magnitude $\eta_0$.  We
assume isothermality and adopt an ideal gas equation of state with
constant temperature, T, specified by our choice of the sound speed,
$c_s$ (see below).  By adopting an isothermal equation of state we
assume that all thermal energy generated through the dissipation of
kinetic and magnetic energies is immediately and efficiently radiated
away and that there is an inflow of heat to adiabatically expanding
regions (such as the buoyant magnetic flux tubes described in \S3
below).  Without a more sophisticated treatment of the thermodynamics
of the gas, we cannot directly measure temperatures produced in our
dynamical models (although by equating the energy dissipation rate
measured in our simulations with an optically thin cooling rate, we
can {\em estimate} equilibrium temperatures that might be reached in
the corona, see \S4).  Moreover, if the vertical structure of the disk
is significantly altered due to heating by the MRI, the rate of energy
transport via buoyancy, and therefore the properties of the corona,
may be affected.  Also, we note that the assumption of isothermal
evolution may increase the buoyancy of the flux tubes (see Torkelsson
\markcite{r}1993) although this effect has not been studied in three
dimensions.  However, dynamical models which include a realistic
treatment of radiation transport within an optically thick midplane
and optically thin upper layers are beyond the current investigation.

We note that the current (the second term on the RHS in equation [2])
has been modified to include Maxwell's displacement current (e.g.
Jackson \markcite{r}1975); see the appendix for the definition of
${\bf J}$ as it is used here.  Physically, the displacement current
limits the speed of propagation of electromagnetic waves to the
constant $c$.  With the usual assumptions of ideal MHD, it can be
dropped from the system.  We keep it here to limit the propagation
velocity of Alfv\'en waves to a constant $c = c_{lim}$, where we set
$c_{lim}$ to be a value much less than the speed of light, but much
larger than any relevant propagation speed within the disk.  We
further discuss the implications of including the displacement current
in the evolution equations, and describe a technique for constructing
a finite-difference approximation to the equation of motion [2]
including it, in the appendix.

We use an initial equilibrium state consisting of a
density distribution which is constant on horizontal planes and follows
a Gaussian profile in the vertical coordinate,
\begin{equation}
\rho(x,y,z)=\rho_o e^{-z^2/H_z^2},
\end{equation}
where $H_z^2=2 c_s^2/\Omega^2$ is the thermal scale height.  Our disk
is thin; i.e., $H_z/R = 0.01$.  Initially all components of the
velocity are zero except $v_{y} = \Omega x$.  The choice of $\Omega$
is arbitrary in the shearing box formalism; for comparison with SHGB,
we choose $\Omega=10^{-3}$ and $2 c_s^2 = 10^{-6}$ (which yields
$H_z=1$).  Adopting $\rho_o = 1$, our initial pressure at the disk
midplane is thus $P_o=5 \times 10^{-7}$.  To seed the MRI, random
perturbations with amplitudes of order $0.001 c_s$ are added to the
velocities.

We study the evolution of three different initial magnetic field
geometries: 1) a purely toroidal field, 2) a Zero Net Z (ZNZ) field,
and 3) a uniform $B_z$ field.  The purely toroidal field is
initialized only within $|z| \le 2 H_z$ such that $\beta = 25$ is
constant within this region.  The ZNZ field is generated by
finite-differencing the following vector potential:
\begin{equation}
 A_x = \left\{ \begin{array}
                    {r@{\quad:\quad}l}
        \sqrt{2 P_o/\beta(0)} \cos(2 \pi x) \cos(y) & |z| \le 2 H_z \\ 
        \sqrt{2 P_o/\beta(0)} \cos(2 \pi x) \cos(y) e^{-(|z|-2 H_z)^4} & 2 H_z < |z| < 4 H_z \\
        0 & |z| \ge 4 H_z
                 \end{array} \right.   
\end{equation}
where $\beta(0)=25$ is the plasma parameter at the disk midplane.  This
vector potential generates a field which has only a vertical component
$B_{z} \propto \sin (2\pi x) \cos y$ within two scale heights, and
which is closed via loops between 2 and 4 scale heights.  This
configuration has  a null mean within $\frac{+}{}$2 $H_z$.  Thus,
within two scale heights, this geometry is equivalent to the ZNZ field
studied extensively by SHGB.  Finally, the uniform $\hat{z}$ field is
a constant field initialized over the whole grid, with $\beta(0)=25$.

Our simulations are computed using the 3D version of the ZEUS code
(Stone \& Norman \markcite{r14}1992a,\markcite{r15} 1992b).  The Ohmic
resistivity term is differenced using the method described in Fleming
et. al. (\markcite{r}1999).  These authors found that $Re_{M} \equiv
H_{z}c_{s}/\eta \leq 10^{4}$ ultimately quenches the MRI after many
orbital times if the initial field configuration has a zero mean.  In
order to explicitly measure the reconnection and resistive heating
rate in the corona, we have computed several models with $Re_M =
20,000$, which, for the numerical resolutions used here, is large
enough to resolve the resistive lengths but small enough not to
completely damp the turbulence driven by the MRI.

Our simulations consist of a grid of size $1 H_z \times 10 H_z \times
10 H_z$, with a range of $-0.5 H_z$ to $0.5 H_z$ in the $\hat{x}$
direction, 0 to $10 H_z$ in the $\hat{y}$ direction, and $-5 H_z$ to
$5H_z$ in the $\hat{z}$ direction.  We have also computed a model in
which the horizontal extent of the box is increased by a factor of
two.  Our standard resolution is $ 32 \times 64 \times 128$ grid
zones.  We partition the zones in the vertical direction such that
there are 64 zones in the region $|z| \le 2 H_z$; thus, inside this
region our resolution is the same as SHGB.  Above $|z|=2$, the grid
zones in the vertical direction increase logarithmically.  We note
that we have run one simulation with a grid which is nonuniform over
the entire vertical domain to test that our choice of nonuniform zone
spacing above $\pm 2 H_z$ does not affect the results reported here
(particularly that it does not affect the vertical structure described
below).  There were no qualitative differences between the test
simulation and our fiducial model (some quantitative differences did
develop within $\pm 2 H_z$ due to differences in resolution).  We have
also run some higher resolution simulations to test the effect of
resolution on our results.  Such simulations are computationally
intensive, prohibiting very high resolution runs.

In the $\hat{x}$ and $\hat{y}$ directions, we adopt the periodic
shearing sheet boundary conditions described in HGB1.  In the
$\hat{z}$ direction, we have implemented free (outflow) boundary
conditions which allow mass, energy, momentum, and magnetic field to
leave the grid.  As described in Stone \& Norman (1992a; 1992b), such
outflow boundary conditions in the ZEUS code are implemented by using
a constant (zero slope) projection of all the dynamical variables into
the boundary zones. 

\section{Results}

Table 1 lists the properties of the simulations discussed in this
paper.  Column 1 gives the identifier for each run; column 2 gives the
value of $\beta$ at the disk midplane; column 3 indicates the value of
$Re_M$; and column 4 gives the value of $c_{lim}$ used in the
displacement current.  Columns 5 and 6 list our numerical resolution
and the ratio of the critical wavelength, $\lambda_c$ (calculated using
equation 15 of HGB1), at the midplane to the grid size, which is given
by $\Delta y$ for the toroidal field runs and by $\Delta z$ for the
$B_z$ runs.  When this ratio is greater than 5, the fastest growing
wavelengths are well resolved (see HGB1).  Column 7 indicates the
total time of the simulation in orbits.  Columns 8, 9, and 10 give the
time and space averaged values of the magnetic energy and the
azimuthal ($\hat{y}$) components of the Maxwell and Reynolds stress
for the regions $|z| \le 2 H_z$ (hereafter referred to as the
``disk'') and $|z| > 2 H_z$ (hereafter referred to as the ``corona'');
these quantities have all been normalized to the initial pressure at
the midplane.  If we adopt the Shakura-Sunyaev $\alpha$ convention,
the time averaged value of $\alpha$ for each simulation is simply the
sum of columns 9 and 10.

\subsection{Toroidal Field Simulations}

We have performed toroidal field simulations with and without explicit
resistivity (the runs with resistivity have been performed at two
different resolutions); these are listed in Table 1 with the prefix
``BY.''  Our fiducial run (BY1) is a standard resolution, nonresistive
run with $\beta$ = 25 in the magnetized region ($|z| \le 2 H_z$). This
simulation has been run for 50 orbital times.  The critical wavelength
for this run corresponds to 8.31$\Delta$y; thus initially it is well
resolved.

Figure 1 is a space-time plot showing the variation of horizontally
averaged magnetic (top panel) and kinetic (bottom panel) energies as a
function of vertical height and orbital time.  The plot is constructed
from the two dimensional function $F(z,t)$ by averaging each quantity
$f$ in the horizontal plane:
\begin{equation}
F(z,t)=\frac{\int\int f(x,y,z,t)dx dy}{\int\int dx dy}.
\end{equation}
The plot of kinetic energy clearly shows the development of strong
fluctuations associated with turbulence in the disk core after
approximately 4-6 orbits driven by the nonlinear evolution of the MRI.
We note that the growth rate and saturation time of the toroidal modes
of the MRI observed here are in agreement with the toroidal field
simulations of HGB1 (c.f. their figure 8) and SHGB (c.f. their figure
7).  From the plot of the magnetic energy, periodic rise of buoyant
magnetic field from the disk core over a few orbital times is clearly
evident.  The magnetic energy in buoyantly rising structures reaches a
maximum at approximately 2.5 $H_z$ and then decreases.  Thus, buoyancy
appears to add both magnetic flux and heat (through the dissipation of
magnetic energy) into the region $z > 2.5 H_z$.  Since kinetic energy
is dominated by the mass distribution, comparison of the spacetime
plots of the kinetic and magnetic energies indicate that the buoyantly
rising field carries a small amount of mass into the coronal region.
Only a small fraction of the mass and magnetic energy escapes the grid
through the vertical boundaries; most remains in the corona.  At the
end of the simulation, the value of $\beta$ at the midplane yields a
ratio of $\lambda_c/ \Delta y$ = 6.14; thus, the fastest growing
wavelengths are still resolved. 

To illustrate the vertical structure produced by the MRI, Figure 2
plots horizontally averaged values of the density, $\beta$, kinetic
and magnetic energies, and the azimuthal components of the Maxwell and
Reynolds stress averaged over orbits 25 through 50, after the
saturation of the MRI.  We also list in Table 2 the volume-averaged
values of each component of the magnetic and kinetic energies, the
off-diagonal components of the Maxwell stress, the azimuthal component
of Reynolds stress, and the density in the disk and coronal regions
averaged over the same times.  The graph of density shows that it is
still roughly Gaussian.  Flaring occurs in the wings due to the
transport of mass into the corona via buoyant flux (from Figure 1 the
expelling of fluid from the disk into the coronal region is periodic
and correlated with the rise of magnetic flux). However, the inner
regions of the disk remain almost unchanged; the total mass in the
corona is only 1.03 \% of the total mass within $2 H_z$.  Table 2
confirms that the coronal densities are two orders of magnitude less
than in the disk midplane, and indicates that there are significant
fluctuations in coronal region, with $\langle \delta \rho^2
\rangle^{1/2}/\langle \rho \rangle \approx 0.37$.  The plasma $\beta$
is greater than one inside the disk and increases to nearly fifty at
the midplane, roughly twice its original value.  This means that the
core of the disk remains unstable to the MRI throughout the
simulation.  Above $\pm 2 H_z$, $\beta$ is $\lesssim$ 0.7, confirming
that a magnetically dominated corona has indeed been produced.  Above
$\pm 2.5 H_z$, $\beta$ becomes small enough that $\lambda_c$ exceeds
$10 H_z$ (the domain size in the azimuthal direction); this is why
this region is stable to the MRI.  Overall, $\beta$ varies by more
than three orders of magnitude from the corona to the disk midplane.
The magnetic and kinetic energy plots also show that the corona is
magnetically dominated.  The magnetic energy in the coronal region
exceeds the kinetic by more than an order of magnitude.  Note that the
magnetic energy in the corona has increased from zero to 0.042 $P_o$
due to the rise of flux from the disk.  It is approximately constant
within the disk core.  The magnetic energy peaks at roughly $\pm 2
H_z$; this same behavior was observed by SHGB (a further indication
that the vertical magnetic energy profile described by these authors
was not entirely a product of the periodic vertical boundary condition
they used).  The scale height of $B^2/(8 \pi)$ is roughly twice the
density scale height.  Referring to Table 2, we see that the azimuthal
component of magnetic energy strongly dominates in both the disk and
coronal regions.  The radial component of the kinetic energy is about
a factor of two greater than the vertical and azimuthal components in
the disk.  In the corona, however, $\rho \delta v_y^2 / 2 P_o$, which
measures the energy in the velocity fluctuations, is nearly two orders
of magnitude greater than the other components.
 
In order to ensure the vertical profiles and volume averaged values
shown in Figure 2 and Table 2 respectively are not influenced by the
time period over which they were averaged, we have reconstructed them
with data which has been time-averaged over orbits 15 through 50 (note
that this period includes the strong transient peak in magnetic energy
associated with the initial growth of the toroidal field MRI evident
around orbit 10 in Figure 1) and over orbits 35 through 50 (beginning
the averages well after the time when this initial peak has faded).
We find that the time averages which begin at orbit 15 (see columns
8,9, \& 10 of Table 1) and those which begin at orbit 35 lead to
nearly the same vertical profiles and volume-averaged values as
reported in Figure 2 and Table 2, indicating these values are not
dominated by transients associated with any particular period of
evolution.  Moreover, in $\S 3.1.2$, we show that higher resolution
simulations produce similar profiles, and in $\S 3.2$ we show that
other field geometries also lead to similar profiles.  This suggests
the vertical structure reported here for our fiducial model is a
general outcome of the evolution of the MRI in highly-stratified
disks.

From Figure 2, we see that the average of the azimuthal component of
the Maxwell stress within $\pm$ 2 $H_z$ is a factor of 4.05 greater
than the azimuthal component of the Reynolds stress, consistent with
the toroidal field simulations of SHGB.  This indicates that the
turbulence and angular momentum transport within the disk are
magnetically dominated.  The profile of the Reynolds stress component
is dominated by the density (although it is slightly flatter within
the disk region), while that of the Maxwell stress shows a very
different pattern.  Note that the azimuthal component of the Maxwell
stress is fairly constant within the disk region, rising only slightly
(less than a factor of 2) at the disk midplane and edges ($\pm 2
H_z$).  This means that $\alpha$ is fairly constant in the disk; our
simulations yield an average value of 0.027 for the $\alpha$ parameter
within the disk core.  Again, we note that we find approximately the
same value of $\alpha$ ($\lesssim 0.03$) whether or not we include the
initial period when the MRI is saturating when we perform the temporal
average.  Thus, even though magnetic energy decreases slightly toward
the end of the simulation (see figure 5), significant angular momentum
transport continues to take place.

In the corona, we find that the azimuthal components of the Maxwell
and Reynolds stresses drop sharply by 1-2 orders of magnitude over
only 3 $H_z$.  The dramatic decrease in stress as we move into the
coronal region indicates both that the transport rate has dropped and
that the flow is much less turbulent here.  The average value of the
azimuthal component of the Maxwell stress in the coronal region is a
factor of 5.0 smaller than in the disk (see Table 2), indicating the
magnetic field is much less tangled.  We find that the profile of the
combined azimuthal stress as a function of vertical height is
approximately flat within $\frac{+}{} 2 H_z$ and declines as
$e^{-z^2}$ above $\frac{+}{} 2 H_z$.  Therefore, the azimuthal
component of the combined stress does not follow the density profile
within the disk core.

Figure 3 is a space-time plot of the off-diagonal components of the
Maxwell stress tensor, as well as the azimuthal component of the
Reynolds stress tensor.  Note that all components except the $B_xB_z$
component of the Maxwell stress are largest in the region $|z| \le 2
H_z$, confirming that the disk flow is turbulent while the coronal
flow is smoother.  However, although the magnitudes of the stresses
are much lower in the corona than in the disk, both regions are
characterized by large fluctuations ($\langle \delta f^2
\rangle^{1/2}/\langle f \rangle \approx 1.0$ (10.0 for the $B_xB_z$
and the $B_yB_z$ components) over the whole grid - see Table 2).  The
$B_xB_z$ component peaks between 2-3 $H_z$, just above the core of the
disk, in the region where the magnetic energy begins to dissipate.
$B_z$ has a local maximum in this region as well, indicating that the
field geometry is most strongly bent upwards by buoyancy between 2-3
$H_z$.  The azimuthal component of the Reynolds stress confirms that
the disk remains turbulent throughout the simulation.  Diagonal
streaks in the azimuthal components of the Maxwell and Reynolds
stresses are clearly associated with the rising magnetic energy.
Peaks in both (but particularly in the azimuthal component of the
Maxwell stress) can be seen in the rising flux tubes.  The $B_xB_z$
and $B_yB_z$ terms are two (one) orders of magnitude smaller than the
azimuthal component of the Maxwell (Reynolds) stress (see Table 2) and
show a highly sporadic pattern.  This suggests that pressure is a more
dominant force than the off diagonal components of the stress in
leading to disk flaring.  The $B_yB_z$ term is of particular interest
as it is associated with the formation of a magnetocentrifugal wind.
Our fiducial simulation produces an average value of 3.71 $\times
10^{-4} P_o$ for $B_yB_z$ above $\frac{+}{} 2 H_z$, suggesting that
only a very weak, unsteady wind is produced.  We note, however, that
the use of shearing sheet boundary conditions in the radial direction
enforces a Keplerian rotation profile throughout the entire grid, even
in the upper regions where the plasma is magnetically dominated.  This
may strongly affect the formation of a magnetorotational wind.  Across
the outer vertical boundary, average mass flux $\dot{M} = 2.21 \times
10^{-4} \sigma_{d} c_s$, where $\sigma_{d}$ is the surface density at
the disk ``surface,'' taken to be $\pm 2 H_z$.

Above $\pm 2 H_z$, the azimuthal components of both Maxwell and
Reynolds stresses drop sharply, decreasing by more than an order of
magnitude.  The low value of the Reynolds term in the coronal region
is, however, more a reflection of the low coronal densities than an
indication of smooth, quiescent flow (c.f.  Table 2 which indicates
that $\rho \delta v_y^2 \sim 0.4-0.9 P_o$ in the corona).  The coronal
region is, in fact, in a highly dynamic state dominated by shocks.
Figure 4 shows the vertical profile of the horizontally averaged rms
velocity dispersion normalized by the sound speed, i.e.,
\begin{equation}
 \langle \frac{(\Delta v)^2}{c_s^2} \rangle^{1/2} = 
\sqrt{\frac{\int\int (\delta v_x)^2 + (\delta v_y)^2 + (\delta v_z)^2 dx
dy}{c_s^2 \int\int dx dy}}.
\end{equation}
The profile shown in Figure 4 has been averaged over snapshots taken
at seven equally spaced orbital times, between orbits 25 and 50.  The
core of the disk is dominated by subsonic turbulence.  The velocities
become nearly supersonic above roughly 3 $H_z$, the height at which
the buoyant magnetic energy dissipates.  In this region, shocks with
Mach numbers as high as 2.9 are observed.  These high velocities form
as acoustic waves generated by the MRI carry part of the energy
generated in the disk into the corona, steepening into shocks and
dissipating there.  The corona is not turbulent, but is highly
dynamic.

It is interesting to ask whether largescale magnetic field structure
is generated in the corona as it becomes magnetized.  Figure 5 is a
plot of the relative strengths of each component of the field energy.
With the onset of the MRI, the energy in all three field components
increases.  However, it is clear that within both the disk and the
corona $B_y^2$ dominates; it is more than an order of magnitude
stronger than either of the other two components in both regimes.
This is not surprising since differential rotation favors the
production of $B_y$.  In the disk we find an ordering of $B_y^2
\approx 15 B_x^2$ and $B_x^2 \approx 2 B_z^2$; this is consistent with
previous three-dimensional studies of the MRI (e.g., HGB1, HGB2, and
SHGB).  Interestingly, the energy in the toroidal field is roughly
equal in the disk and corona by the end of the run.  This is due both
to the efficiency of buoyancy in magnetizing the coronal region and to
our shearing sheet boundary conditions which force Keplerian rotation
and may thus be favoring the production of $B_y$.  After only 10
orbital times the corona is fully magnetized; its configuration is
roughly that of a well ordered, toroidal magnetic field.  In the
corona, $B_x^2$ and $B_z^2$ are approximately equal.  Differential
vertical velocities therefore increase $B_z$ somewhat in the coronal
region but are not enough to produce a strong $B_z$ field here; the
toroidal component of the field is larger than the poloidal by more
than an order of magnitude.  This differs slightly from the results of
SHGB, who report an ordering of $B_x \approx 3 B_z$ in both the corona
and disk regions, suggesting the increase in the vertical domain of
our simulations has allowed buoyant motions to develop more fully.  We
see largescale toroidal field configurations in both the corona and
the disk, with small scale ``wiggles'' in the $\hat{x}$ and $\hat{z}$
directions.  The slight secular decrease in the total disk magnetic
energy is consistent with previous toroidal field studies; simulations
on timescales longer than is possible here show the field never
completely dies away.

\subsubsection{Dissipation of Energy}

Turning now to the resistive equivalent of our fiducial run (BYR1), we
quantify the roles of dissipation and transport by examining the time
rate of change of magnetic energy in the disk and coronal regimes.  We
use the fiducial resistive run for this analysis because we are able
to explicitly measure resistive dissipation rates.  In each of these
volumes, magnetic energy is lost through a combination of the Poynting
flux out of the volume and conversion into other forms of energy.
Similarly, it is generated by conversion from other forms of energy
(e.g. kinetic energy associated with the shear by the MRI) and the
Poynting flux into the volume.  If we let $S_V$ denote the net
generation of magnetic energy (embodying field generation by the MRI
and energy loss due to conversion into other forms of energy), then we
can write an equation of magnetic energy conservation as follows:
\begin{equation}
S_V = \frac{\partial \int_V (B^2/8 \pi) dV}{\partial t} - \oint_S (B^2/4 \pi) {\bf v} \cdot d {\bf S} + \frac{1}{4 \pi} \oint_S ({\bf B} \cdot {\bf v}) {\bf B} \cdot d {\bf S} - \int_V \eta J^2 dV,
\end{equation}
where the last term on the right indicates losses due to resistive
heating and the second and third terms represent the Poynting flux out
of the volume.  We can characterize the dissipation of energy
throughout the volume by comparing the relative strengths of each of
these components in the disk and coronal regimes.  We will also
compare with the heating rate due to compression and artificial
viscosity, $R_{heat}$, given by:
\begin{equation}
R_{heat} = \int_V (P \nabla \cdot {\bf v}) dV + \int_V ({\bf q}
\cdot \nabla {\bf v}) dV,
\end{equation}
where the components of ${\bf q}$, given by
\begin{equation}
q_i = 4 \rho (d v_i/d x_i)^2,
\end{equation}
are the artificial viscous pressure used in the ZEUS code to capture
shocks (thus the viscous heating as we have defined it includes shock
heating), and we will compare with the flux of kinetic energy,
$F_{kin}$:
\begin{equation}
F_{kin} = (1/2) \oint_S (\rho v^2) {\bf v} \cdot d {\bf S}.
\end{equation}
We will report each quantity in units normalized by the initial
pressure, $P_o$, per orbit.  Time averages are computed for orbits
15-50, after saturation of the MRI.  In addition, we have plotted the
instantaneous net magnetic energy generation rate ($S_V$) in Figure 6
as a function of orbital time in the disk (left panel) and coronal
(right panel) regions.  In each plot, $S_V$ has been averaged
spatially within $\pm$ 2 $H_z$ (left panel) and above $\pm$ 2 $H_z$
(right panel) and normalized by the initial pressure, $P_o$.

In the disk, $S_V$ has a net positive time-averaged value of 0.10, the
average magnetic energy generation rate is $\approx -0.0005$, and the
average Poynting flux out of the volume is 0.10.  Resistive heating is
very small, with an average value of 0.0023.  In the disk, the
evolution of the $S_V$ term (see Figure 6a) is dominated by largescale
($\delta S_V \gg \langle S_V \rangle $) fluctuations; these are
related to the production of field via the MRI and to the loss of flux
via buoyancy.  The net positive time average value of $S_V$ indicates
there is a local increase in magnetic energy within the disk.  This
agrees well with the results of SHGB (see their figure 11); they also
found the evolution of $S_V$ to be dominated by fluctuations and
report a net positive value within the disk core.  (The major
difference is that we do not see their initial spike in $S_V$; this is
because we start with a toroidal magnetic field instead of a poloidal
field).  Our average value is more than an order of magnitude larger
than that reported by SHGB, probably due to the fact that we measure
$S_V$ at two scale heights at the boundary between the turbulent core
and stable corona, where buoyancy fluxes are a maximum (SHGB reported
values measured at only one scale height).

Comparing generation and loss rates in the disk, we find that energy
generation via viscous heating dominates.  Loss of magnetic energy due
to resistive heating is quite low, representing only 0.71 \% of the
viscous heating rate.  The average Poynting flux out of the disk
represents 30.9 \% of the viscous heating rate while the flux of
kinetic energy is only 5.03 \% and has a net negative average (which
represents flow into, not out of, the disk).  Thus, the rate of energy
loss as flux tubes rise out of the disk is about one quarter of the
total disk energy generation rate.  This shows that the MRI
regenerates field quickly enough to ensure a net increase in magnetic
energy and to sustain turbulence and angular momentum transport for
the duration of the run in spite of significant field loss due to
buoyancy.  We also note that at two scale heights, the amount of
magnetic energy lost via buoyancy is more significant than the value
reported by SHGB at one scale height.

In the corona, the time average value for $S_V$ is -0.012.  The
magnetic energy generation rate has an average of 0.0002, the net
Poynting flux into the volume is 0.011 and the average resistive
heating is 0.00017.  Comparing these values with those in the disk, we
see that the situation in the corona is very different.  The net
Poynting flux rate is positive which means that more magnetic energy
enters the corona than leaves it.  However, the mean energy changes
only slightly with time and the net value of the $S_V$ term is
negative.  This indicates there is a local decrease in magnetic energy
in the coronal region despite the net influx of field.  We therefore
conclude that the excess magnetic energy transported into the coronal
region through the $z= \pm 2 H_z$ boundaries is dissipated
there.  Looking at Figure 6b, we see that $S_V$ shows large
fluctuations; these correlate very well with the rise of flux tubes
into the volume.  The correlation of the net flux into the corona with
the fluctuations in the net magnetic energy generation rate (the $S_V$
term) indicates that transport of energy dominates local processes in
the corona.  Therefore, although there is a net transport of MRI
generated flux into the corona, this region remains stable: no
magnetic energy generation occurs and the net magnetic energy
transported into the corona is dissipated there.  We note that SHGB
also report a net loss of field in the coronal region, although, as
before, their net value for $S_V$ is an order of magnitude smaller.

The energy in the net transported magnetic field is 60.1 \% of the
viscous heating in the coronal region; magnetic energy is thus the
dominant form of heating here.  However, it represents only 3.4 \% of
the viscous heating in the disk; therefore, it is only a small
fraction of the overall energy budget.  Resistive dissipation in the
corona is surprisingly small.  This is most likely due to the fact
that the field structure is predominantly toroidal in both the disk
and corona (as discussed above).  Therefore, it is already fairly
ordered as it rises into the corona.  This also means that loss of
magnetic energy in the coronal region is due more to its conversion
into other forms of energy (e.g., kinetic energy) than to reconnection
(as measured by $\eta_o$). The flux of kinetic energy is comparable in
magnitude to the Poynting flux, although its average value is
negative, indicating a net flow both back into the disk region and out
of the upper boundary of the simulation (roughly 10 \% of the kinetic
flux escapes through the upper boundary). Thus, energy enters the
corona mostly in the form of magnetic energy.  We conclude that the
heating of the corona occurs via the conversion of magnetic energy
into kinetic energy which is then dissipated by the viscosity.

\subsubsection{Effect of Resistivity and Resolution}

The evolution described above is typical of all toroidal field runs.
The resistive runs show growth due to the MRI at slightly later
orbital times, but otherwise follow the same qualitative evolution.
As we increase the resistivity (i.e. from $R_M = 20,000$ in run BYR1 to
$R_M = 10,000$ in run BYR2), the effect increases (i.e., growth occurs at
increasingly later orbital times), but the general evolution does not
change.  The delay in growth with the addition of resistivity occurs
because resistivity tends to diffuse gradients in the field, resulting
in damping of the fastest growing wavelengths (Sano et. al 1998;
Fleming et. al.  \markcite{r23}1999).

Run BYR3 is a high resolution version of the fiducial resistive run
(BYR1); it was run on a grid containing twice the standard resolution.
Again, the qualitative evolution remains the same as that described
above.  Comparing over the same time periods, the post saturation
volume averaged values of magnetic energy increase by a factor of 1.3
in the disk region and 1.8 in the coronal region.  We find the same
ordering in the relative strengths of the individual components of
magnetic energy in both regions regardless of resolution (see Figure
5).  The azimuthal components of both the Reynolds stress and Maxwell
stress increase with resolution by factors of 1.6-1.5 (respectively)
in the disk region.  In the corona, the azimuthal component of the
Reynolds stress remains nearly the same, increasing by only a factor
of 1.2 while that of the Maxwell stress increases by a factor of 1.5.
Thus, doubling the resolution has made important differences to some
components of the energy and stresses, but in all cases this
difference is less than a factor of two.  The increases in magnetic
energy and stress are probably due to an increase in the range of MRI
unstable wavenumbers which are resolved in the simulation.  Overall,
the $\alpha$ parameter increases by a factor of 1.5 in the disk as we
double the resolution.  Most importantly, plotting the equivalent of
Figure 2 for run BYR3, we find the profiles of all four quantities to
be almost identical.  $\beta$ spans the same range in values from the
disk midplane to the corona, and mean magnetic and kinetic energy
values are the same in both regions as are mean values of the
azimuthal components of the Maxwell and Reynolds stresses.  The only
noticeable difference in the graphs is in the magnetic energy and the
azimuthal component of the Maxwell stress; these both dip to lower
values at the disk midplane in the standard resolution runs instead of
remaining approximately constant throughout the disk as they do in the
high resolution run.  Again, this indicates that we may not be
resolving all of the unstable wavelengths in the disk midplane in the
standard resolution simulations.  Shock and resistive heating remain
the same as in the standard resolution runs. Poynting flux also
remains the same over the entire grid while the flux of kinetic energy
increases with resolution by factors of 4.8 in the disk and 2.0 in the
corona.  However, the velocity dispersion, which characterizes the
turbulence of the flow, has the same profile and average values as in
the standard resolution runs (see Figure 4).  We find that the
evolution of $S_V$ in both the disk and the coronal regions is the
same as that described above for the standard resolution runs.  The
average value of $S_V$ in the disk increases by 15\% in the high
resolution run; the average coronal value remains unchanged.  It
appears that our standard resolution simulations capture the
qualitative evolution very well, although they may underestimate some
quantities, such as the $\alpha$ parameter.

We have also varied the initial field setup to ensure that our choice
of $\pm$ 2 $H_z$ as the initially magnetized region is not reflected
in the results.  We tried limiting this initially magnetized region to
$\pm$ 1 $H_z$ as well as holding {\bf B} constant within $\pm$ 2
$H_z$, enforcing $\beta$ =25 at the equatorial plane only; these are
also listed in Table 1 as BY2 and BY3, respectively.  None of these
changes resulted in any significant differences with the general
evolution described above.  Finally, we repeated the fiducial run
doubling the radial domain (run BY4).  The small radial extent of our
simulations may be limiting the off-diagonal components of the stress
tensor and thus the production of magnetocentrifugal winds.  Our small
radial domain size may also be limiting transport rates and fluxes.
However, doubling the radial domain size was not enough to produce any
significant difference.  Magnetic energy and magnetic and kinetic
stresses actually decreased (by factors of 1.1-1.2) in both the disk
and coronal regions when the radial domain was doubled.  This leads to
a slight decrease in the $\alpha$ value for the disk in run BY4 (by a
factor of 1.2) compared to the value in the fiducial run.  The
Poynting flux and flux of kinetic energy remained the same at both the
disk surface ($\pm$ 2 $H_z$) and the outer boundary of the simulation
box.  There was also no change in the net rate of generation of
magnetic energy (the $S_V$ term) in the coronal and disk regions.  We
did not continue increasing the domain size due to the amount of
computational time required.  Clearly, the most fruitful means of
investigating these issues is to perform fully global simulations
which span a large range in radii.

\subsection{Zero Net Z Simulations}

The Zero Net Z field simulations are identified in Table 1 by the
prefix ``ZN.''  Our fiducial simulation (ZN1) is a standard
resolution, nonresistive run with $\beta(0)$=25.  The simulation was
run for 50 orbital times.  The critical wavelength for this run
corresponds to 20.8 $\Delta$z (within the disk midplane); it is well
resolved.  Note that in SHGB this field configuration was the fiducial
model.  In this paper, we adopt toroidal field runs as the fiducial
model both because a toroidal field configuration is physically
motivated by the differential rotation of the disk and because we have
to close $B_z$ fieldlines in the coronal region, producing curved
field lines there.  (This was not an issue for the models presented in
SHGB since they used periodic boundary conditions at $\pm 2H_z$.)  The
primary effect of the curvature of the field lines in the upper layers
of the disk is to create tension forces which cause motions in the low
density regions long before the MRI has saturated in the disk
midplane.

Despite the change in initial magnetic field configuration, the
qualitative evolution of the ZNZ runs is very similar to that of the
toroidal field simulations (described above).  As expected (see HGB1,
HGB2, SHGB), the MRI emerges earlier (after only 2-3 orbits) than in
the toroidal field runs.  Figure 7 is a space-time plot of run ZN1
showing the vertical variation of horizontally averaged magnetic
energy (top panel), the azimuthal component of Maxwell stress
(2$^{nd}$ panel), kinetic energy (3$^{rd}$ panel), and the azimuthal
component of Reynolds stress (bottom panel) versus orbital time.  The
MRI develops initially in the upper layers of the disk.  By three
orbits, however, the dominant modes are those at the disk midplane.
As in the toroidal field runs, the MRI increases the magnetic energy
within the disk region, and the magnetic energy rises buoyantly into
the corona.  Strongly magnetized regions rise with approximately the
same slope (as measured in the space-time plots) as in the toroidal
field runs and saturate at 3 $H_z$, creating a magnetized corona.  As
in the toroidal field simulations, this process is periodic on a short
(3-4 orbit) timescale.  The major difference between run ZN1 and the
toroidal field runs is that here the instability appears to die away
at the midplane at the end of the run; the average value of $\beta$ at
the midplane at late times yields a ratio of $\lambda_c/\Delta z =
5.9$, indicating that the critical wavelengths are just barely
resolved.

The main quantitative differences between the ZNZ and the toroidal
field simulations lie in the magnetic quantities, which are uniformly
smaller in the ZNZ runs.  This is most likely due to the fact that the
critical wavelengths are not well resolved in the midplane by the end
of the ZNZ simulations.  Figure 8 shows the horizontally averaged
values of the density, $\beta$, kinetic and magnetic energies, and the
azimuthal components of the Maxwell and Reynolds stresses averaged
over orbits 10-50, after saturation of the MRI.  Comparing Figure 8
with Figure 2, we see that the range in density is approximately the
same for both field configurations.  The average coronal values of
$\beta$ are about the same, although $\beta$ is 6 times larger in the
midplane in the ZNZ run, so that it spans slightly more than 4 decades
over the vertical extent of the grid.  The range in the azimuthal
component of the Maxwell stress is about the same in both runs,
although the absolute values are smaller in the ZNZ field runs.
Probably the most notable feature of the profiles is a sharp dip at
the midplane in both the magnetic energy and the azimuthal component
of Maxwell stress.  This is largely due to the strong transient burst
of magnetic energy that occurs around 2 $H_z$ between orbits 30 and 40
(see Fig. 7).  The kinetic energy follows the density profile closely
and has about the same range as in the toroidal field runs.  The
azimuthal component of the Reynolds stress drops more sharply in the
disk surface layers and corona compared to the toroidal field runs,
although it spans approximately the same range.  The disk $\alpha$
parameter for this field configuration is about 0.01, which is about a
factor of 3 smaller than the value found in the toroidal field runs.
We assign no significance to the dip in the magnetic quantities at the
midplane (since it is caused by a transient feature); neglecting the
dip, the profile of the stress for these runs follows the form we
found for the toroidal field runs; i.e., it is approximately flat
within the disk core and is proportional to the density in the corona.

In spite of its initially vertical structure, the field within the
disk becomes predominantly toroidal due to differential rotation
before beginning its buoyant rise and is thus toroidal in the coronal
region as well (see Figure 9).  Comparing Tables 2 and 3, we find the
same relative ordering of the components of magnetic energy in both
the disk and coronal regions as we did in the toroidal runs (see
Figure 5 for comparison).  The ordering of the kinetic energies in
both the disk and coronal regions is also the same as in the toroidal
field runs.  As before, all quantities show large fluctuations.

We find that energies and stresses are generally slightly lower in the
ZNZ simulations than in the toroidal runs.  Magnetic energies in the
disk are factors of 3-4 lower in the ZNZ runs, and in the corona
factors of 2-3 lower.  Surprisingly, even the energy in the $\hat{z}$
component of the field are lower in the ZNZ field simulations, in
spite of the initial vertical field configuration.  Kinetic energies
are lower by factors of 2-3 in the disk and 1-2 in the corona.
Magnetic and kinetic stresses are also lower, with ZNZ values
averaging 4 (2) times smaller in the disk for the magnetic (kinetic)
quantities and 3 times smaller in the coronal region for both magnetic
and kinetic quantities.  Two notable exceptions are the $-B_xB_z$
term, which is nearly identical in the coronal region to the toroidal
field value, and the coronal value of the $-B_yB_z$ term, which is
almost a factor of 10 smaller for the ZNZ field simulations.  We find
an $\alpha$ parameter of .0079 for our fiducial ZNZ run; this about a
factor of 3 smaller than that cited for the toroidal field simulations
($\alpha = 0.027$).  The $B_yB_z$ and $B_xB_z$ components of the
stress tensor show a much stronger correlation with the rising
magnetic flux tubes described above, indicating this field
configuration may be more favorable to the production of ordered
stress patterns which can fuel magnetocentrifugal winds; however,
their time averaged values are still very low (7.2 $\times 10^{-5}$
$P_o$ for the $B_yB_z$ component) in this region.  Magnetic fluxes are
roughly a factor of 3 smaller than those measured in the toroidal
field runs; kinetic fluxes are factors of 2-3 smaller.  However the
ratios of flux to energy for both magnetic and kinetic quantities are
approximately the same; this may explain why the coronal field
geometry at saturation is identical to that seen in the toroidal field
runs.  Overall, there are surprisingly few qualitative differences
between simulations run with ZNZ and toroidal initial field
configurations, although most quantities are slightly lower in the ZNZ
runs.

We note that we repeated run ZN1 with resistivity (run ZNR1), using
$R_M = 20 K$ as in the fiducial resistive toroidal field simulation.
We find that the effects of resistivity on the ZNZ simulations is the
same as that discussed above for the toroidal field simulations.  (See
Fleming et. al. (\markcite{r23}1999) for a more complete discussion of
the effect of resistivity on different initial field configurations.)

\subsection{Pure $B_z$ Simulations}

The evolution of a pure $B_z$ field geometry is interesting because it
is relevant to a circumstellar disk threaded by a stellar magnetic
field; i.e., a disk with a net vertical flux.  Although far from the
central star the stellar field is likely to be dipolar, in the local
approximation adopted here we represent the stellar field as a uniform
strength $B_z$ field.  These simulations are listed in Table 1 with
the identifier "BZ."  Our fiducial simulation (BZ1) is a standard
resolution, nonresistive run with $\beta(0)$ = 25.  This simulation
has been run for 15 orbital times.  However, we will only discuss the
evolution up to 5 orbital times; beyond this point, the disk geometry
is destroyed and the grid becomes magnetically dominated everywhere.
The shearing sheet boundary conditions, which enforce Keplerian
rotation everywhere, are then no longer appropriate for following the
further evolution.

Figure 10 is a space-time plot showing the vertical variation of
horizontally averaged magnetic energy (top panel), the azimuthal
component of Maxwell stress (2$^{nd}$ panel), density (3$^{rd}$
panel), and the azimuthal component of Reynolds stress (bottom panel)
versus orbital time (along the $\hat{x}$ axis).  It is immediately
obvious from a comparison of Figure 10 with Figures 1, 3, and 7 that
the evolution of a uniform $B_z$ field is completely different from
that of the other field geometries discussed in this paper.  The
difference is due to the development of the axisymmetric phase of the
MRI known as the channel solution (Hawley \& Balbus 1991; Goodman \&
Xu 1994).  The channel solution develops at about 3 orbits; it is
characterized by exponential growth of the magnetic field which causes
the disk to separate vertically into horizontal planes (axisymmetric
``channels'' - see panel 3 of Figure 10).  Note that the magnetic
energy and the azimuthal components of the Maxwell stress are large in
the regions between the channels; these rise due to a combination of
magnetic buoyancy and pressure.  The uppermost of these channels rise
out of the simulation box in less than 1 orbital time (these can just
barely be seen rising from the low density surface layers of the
disk).  The rise of the channels which originate closer to the
midplane is checked by the onset of the Parker instability, which
causes undulations in the rising density channels.  Mass slides along
the bends in the channels back towards the midplane, reforming a
semblance of a disk structure which oscillates about the midplane
before reaching an equilibrium state.  The simulation is magnetically
dominated everywhere by the end of the run.

Figure 11 is a snapshot of the vertical structure at 5 orbits, after
the disk has been reformed.  The quantities in Figure 11 have been
averaged spatially in the horizontal directions.  The plot of density
shows that the bulk of the disk is at the midplane, although its
initial symmetric structure has been perturbed.  The plot of $\beta$
indicates that both the disk and the coronal regions are magnetically
dominated; $\beta \lesssim 1.0$ everywhere.  We see from the plot of
magnetic energy that the field has become very large at the midplane;
it is of the same order of magnitude as the kinetic energy in the disk
and is suprathermal over the entire grid.  This is very different from
the toroidal field runs (see Figure 2), where the disk field remained
subthermal throughout the evolution.  The strong magnetic pressure at
the midplane drives the asymmetric density distribution.  The plot of
the azimuthal components of the stresses indicates that the Maxwell
term dominates over the Reynolds term everywhere, and that the Maxwell
term varies by at most a decade over the entire grid.  Thus, after the
disk reforms, matter is highly magnetically dominated in both the disk
and coronal regions.  The field structure is tangled and complicated,
with $B_y^2 \approx B_x^2 \approx B_z^2$ over the entire grid.

We have also completed runs in which the initial density profile is a
function of $x$ (BZ2) and of $x$ \& $z$ (BZ3).  Such density
distributions seed faster growing modes of the Parker instability,
allowing it to curb the violent rise of the buoyant channels more
quickly.  In these simulations, mass loss due to the channels is less;
however, the general evolution is qualitatively similar to that
described above.

We conclude that purely vertical fields quickly lead to a magnetically
dominated disk everywhere.  Simulations beginning with purely vertical
fields were attempted by SHGB, and similar evolution was noted,
however because of the much smaller vertical domain used by these
authors, the buoyant channels would hit the boundary at $\pm 2H_z$
before the Parker instability could grow.  Further simulations of this
initial field geometry will require global models in order to capture
the dynamics of the strongly magnetized regions properly (without
enforcing Keplerian rotation).

\section{Discussion}

In this paper, we have shown that the evolution of a weakly magnetized
stratified disk is characterized by the generation of magnetic field
via the MRI and the loss of magnetic field through the buoyant rise of
flux tubes (as long as the channel solution is not excited).  The
magnetic energy of buoyantly rising field saturates at about 2.5-3.0
$H_z$ and thereafter dissipates, serving to both magnetize and heat
the corona (the region above $\pm 2 H_z$).  The presence of a hot
corona above an accretion disk has important observational
consequences, as discussed in the introduction.  It is therefore
interesting to quantize the dissipation of total energy
($E_{tot}=E_{mag}+E_{kin}+E_{th}$) as a function of height above the
disk and to use this to calculate a thermal equilibrium temperature
profile in the corona assuming the heating rate is balanced locally by
optically thin cooling.  In this section, we will estimate the coronal
temperature distribution for two important cases: 1) a protostellar
disk around a CTTS and 2) a disk around a neutron star, and compare
our results with the coronal temperatures measured by observations.

We first examine the case of a typical protostellar disk.  In order to
convert quantities measured in our simulations into physical units, we
adopt the average CTTS parameters found in Beckwith
et. al. (\markcite{r21}1993) (i.e., mass of disk $\approx$ $0.03
M_{\odot}$; radius of disk $\approx 50 AU$).  We assume our simulation
box is located at $R=3 R_{\ast}$ (the inner edge of the disk) and that
within two scale heights the disk is optically thick and has a
temperature of 1000 K. The total dissipation rate as a function of
vertical height, $D$, we approximate as:
\begin{equation}
D(z) = R_{heat} + (\oint_S \eta J^2 dx dy) \times \Delta z,
\end{equation}
where $R_{heat}$ is as defined in equation [9] (except here we
integrate over a differential volume (corresponding to a horizontal
plane with a height of one grid zone) instead of the entire simulation
domain), and each quantity in the sum is averaged over horizontal
planes at each vertical position in the numerical grid, and also
averaged over time.  We convert the total dissipation rate into an
average per particle heating rate, $\Gamma$, in erg $cm^2$ $s^{-1}$
via
\begin{equation}
\Gamma = D/(\epsilon V \bar{n}^2),
\end{equation}
where $\epsilon$ is a dimensionless heating efficiency and $\bar{n}$
is the average number density of particles within the volume V.  We
find that the shock heating term dominates the other dissipative
components in our simulations; because it tends to be strongly
localized, we set $\epsilon = 10^{-3}$.  To be consistent with our
definition of $D$, the volume V is the planar area $\Delta x \Delta y$
times $dz$, where $dz$ is the height of a vertical grid zone in units
of $H_z$.  To compute the equilibrium temperature, we equate the per
particle heating rate with an optically thin cooling rate, $\Lambda$,
and solve for the temperature $T$.  Using data from run BYR1, averaged
over orbits 15 through 50, we find the profile shown in Figure 12,
where we have used the cooling rate given in figure 2 of Dalgarno \&
McCray (\markcite{r22}1972, hereafter DM) to yield the temperature
scale.  Note that our assumption of a disk temperature at $\pm 2 H_z$
of $T(\pm 2 H_z) \sim 1000 K$ is at least consistent with our result.
In calculating the temperature at this point, we have used the curve
in DM corresponding to the highest degree of ionization.  This seems
reasonable in the low density coronal regions; however, we note that
different assumptions about the degree of ionization at $2 H_z$ would
lead to a different temperature for the same amount of dissipation.

The thermal equilibrium temperature is of order $10^4$ K in the
coronal region.  This is consistent with the warm temperatures Kwan
et. al. (\markcite{r5}1997) need to account for the O~I and Balmer
lines of CTTS.  For comparison, we apply our model to the case of
1548C27, using the stellar parameters given by Najita
et. al. (\markcite{r24}1996).  Asumming we are at their fiducial
radius, we find a temperature $\approx$ 9,000-10,000K for the inner
disk atmospheric temperature.  This is consistent with the temperature
they derive to model the CO overtone emmision in this star.  Since our
simulations are local, we cannot meaningfully compute a radial
temperature profile for the corona to compare to their data.  We
emphasize that the fluctuations in $D$ about the mean value are large
(factors of 4) which implies that the heating rate and peak
temperatures in the corona will be highly time dependent.  We find the
temperature can vary by as much as two orders of magnitude on short
($t \ll t_{orb}$) timescales.  Thus, our simulations show that a
turbulent protostellar accretion disk produces a nonturbulent,
magnetized, warm ($T \sim 10^4$) corona, with short timescale
temperature fluctuations.

We now turn to the case of an accretion disk surrounding a compact
object.  Adopting typical values of 1.0 $M_{\odot}$ for the mass of
the neutron star, $10^8$ cm for a representative disk radius, and
$\dot{M} = 1.6 \times 10^{-9} M_{\odot} yr^{-1}$ as a typical
accretion rate, we can use the formulae in Lovelace
et. al. (\markcite{r25}1995) to derive a fiducial disk mass, velocity,
and time.  These allow us to convert the total dissipation rate into
the per particle heating rate, $\Gamma$, as before.  Using the neutron
star parameters and comparing with the cooling rate given in figure 1
of Raymond et. al. (\markcite{r26}1975), we find that a typical
thermal equilibrium temperature in the corona is of order $10^8$ K at
$3 H_z$, which is in the X-ray range.  Thus, we find the amount of
energy generated by the MRI in the midplane of the disk and
transported into the corona via buoyancy is sufficient to heat the
corona to temperatures inferred from observations for disk coronae
around compact objects.

It is encouraging that the estimates presented in this section are
consistent with temperatures suggested by current observational data
as well as numerical calculations.  However, we caution that there are
many uncertainties in our estimates, most particularly in our choice
of $\epsilon$, which, although motivated by the strong dominance of
shock heating in our simulations, was admittedly ad hoc. If the
heating were less localized, $\epsilon$ could be as high as 1.0, which
would change the temperature range significantly.

\section{Conclusions}

We have performed quasi-global MHD simulations of the dynamical
evolution of vertically stratified accretion disks which demonstrate
the generation of a magnetized corona.  Initially weak magnetic fields
in the core of the disk (below two scale heights) are amplified via
MHD turbulence driven by the MRI.  This field becomes buoyant and
rises out of the disk.  The magnetic energy density associated with
buoyantly rising magnetic field peaks at about 2.5-3.0~$H_z$ and then
dissipates, creating a magnetized, heated coronal region above the
disk core.  We show that this process is highly time-dependent;
throughout our simulations, the accretion disk continues to generate
field via the MRI, which leads to the rise of flux out of the disk and
its subsequent dissipation in the corona.  On long time-averages, the
vertical structure of the disk is a turbulent core which is unstable
to the MRI surrounded by a highly dynamic, magnetically dominated
($\beta \lesssim 10^{-1}$), heated, stable corona. The largescale
field configuration in both the disk and the coronal region is
toroidal.  Our simulations are the first to follow the highly coupled
amplification, buoyant transport, and saturation processes which
control the magnetic field, as well as the angular momentum transport
process which controls the disk, from first principles.

We have characterized the dissipation of energy as a function of
vertical height for these simulations.  We find that local processes
dominate over transport in the disk region.  This leads to heating of
the disk by turbulence and to the continuation of angular momentum
transport via the MRI throughout the simulations.  In the coronal
region, however, transport processes dominate because it is strongly
magnetized ($\beta < 1.0$) and therefore stable to the MRI.  Shocks
with Mach numbers as high as $M \approx 2.9$ are observed in the corona
throughout the evolution; these are produced both by the steepening of
MHD waves associated with the turbulence in the disk as they propagate
into the low density corona, and via mass motions into the corona
itself.  By equating the time-averaged per particle heating rate in
the corona with an optically thin cooling rate, we estimate the
equilibrium temperature of the corona is of order $10^4$ K for a
typical accretion disk around a CTTS, and of order $10^8$ K for a disk
around a typical neutron star, although there is considerable
uncertainty in these estimates.  We note that the temperature can vary
by as much as two orders of magnitude on short ($t \ll t_{orb}$)
timescales.  The functional form of the stress as a function of
vertical height is flat within $\pm 2 H_z$ but falls off proportional
the density (i.e., as a Gaussian) above $\pm 2 H_z$.

We find that these results are general in the sense that they do not
depend on resolution, resistivity (so long as the MRI is not
quenched), nor initial field configuration (as long as the
axisymmetric linear phase (the channel solution) of the MRI is not
excited).  They do depend on the initial field strength (if the field
is strong, $\beta > 1.0$, and the disk becomes MRI stable).  They may
also be affected by the vertical structure of the disk.  We assume an
isothermal profile in all the simulations presented here; adiabatic
disks may not be as strongly stratified and this may affect the
buoyant transport rates.  We are also limited by the use of the local
approximation in the plane of the disk, which enforces Keplerian
rotation everywhere, including in the strongly magnetized corona.
This may suppress the generation of MHD winds.  Extending our models
to be global in all three directions is the natural next step in this
study.

We have also presented simulations in which the axisymmetric linear
phase of the MRI is excited.  Unlike the evolution described above,
these simulations are dominated by the formation of density channels,
which rise quickly out of the disk due to a combination of magnetic
buoyancy and pressure.  The rising density sheets are subject to the
Parker instability, which results in mass sliding back to the midplane
to reform a disk structure.  This highly dynamic evolution may be
relevant to the processes by which a stellar magnetic field can
disrupt an accretion disk within the star-disk interaction region.

We thank Steve Balbus for suggesting improvements to the Alfv\'en
speed limiter algorithm, and John Hawley and the referee (Ulf
Torkelsson) for useful comments.  This work was supported by the NSF
through grant AST95-28299.

\appendix

\section{Large Alfv\'en Speed Limiter via the Displacement Current}

An important aspect of this work is the extension of the simulation
domain to include ten vertical scale heights (five above and five
below the disk midplane).  In time explicit codes such as ZEUS, the
time step is constrained in part by the shortest crossing time of an
Alfv\'{e}n wave across a grid zone.  Thus, low density regions (such
as those three to five $H_z$ above the disk midplane) where the
Alfv\'en speed becomes unmanageably large must be avoided.  In the
current simulations, we are able to work in the low density coronal
region through the addition of the displacement current in the
equation of motion [2], combined with a artificially low value for the
speed of light $c$.  This method was first used by Boris
(\markcite{r16}1970) (see also the review by Brecht
\markcite{r17}1985).

In practice, we set $c=c_{lim}$ (an arbitrary scaling speed) in the
displacement current term in equation [2], which constrains
$v_A/c_{lim}$ to be $\lesssim 1$.  We can thus choose $c_{lim}$ such
that $\Delta x/v_A$ is a reasonable value everywhere in the simulation
domain.  In this work, we choose $c_{lim}$ such that it is always
approximately an order of magnitude greater than the (isothermal)
sound speed; $c_{lim}$ = 5.66 $\times 10^{-3}$.  This ensures that we
allow a good dynamical range in the runs; i.e., the Alfv\'en speed and
the sound speed are allowed to differ significantly.  It also makes
our omission of relativistic terms introduced by the displacement
current reasonable (i.e., $v_A/c_{lim} \lesssim 1$, but in general
$v/c_{lim} \ll 1$, where $v$ represents the other velocities in the
simulation).  We also note that we have tested the effect of this
limiter by varying it over a range of values up to the value we report
in this paper (compare columns 8, 9, \& 10 of Table 1 for runs BY5,
BY6, \& BY1 and ZN2 \& ZN1).  As $c_{lim}$ is increased, the only
effect is an increase in the coronal kinetic energies and stresses.
This is not unexpected, since we are limiting magnetic stresses, and
therefore velocities, in the coronal region through the use of
$c_{lim}$.  The amplitudes of the magnetic energy are not
significantly changed.

Our derivation proceeds as follows.  Consider the momentum equation:
\begin{equation}
\rho \frac{\partial{\bf v}}{\partial t} + \rho {\bf v} \cdot \nabla
{\bf v} = - \nabla P + \frac{1}{c} ({\bf J} \times {\bf B}),
\end{equation}
where vertical gravity and the shearing box rotational terms have been
omitted.  The displacement current enters through the definition of
the current:
\begin{equation}
{\bf J}=\frac{c}{4 \pi}[\nabla \times {\bf B} - \frac{1}{c} \frac{\partial {\bf E}}{\partial t}] + \xi^+ {\bf v},
\end{equation}
where ${\bf E}$ is the electric field and $\xi^+$ is the charge
density. The last term represents the electric field force on the
plasma.  We can self-consistently include it in the definition of the
current because it does not affect the energy balance in the code
(e.g., ${\bf v} \times {\bf B} \sim {\bf E}$, and ${\bf v} \cdot {\bf
E} = 0.0$).  The motivation for choosing this form for ${\bf J}$ will
become apparent below.

Assuming infinite conductivity\footnote{We note that the terms
introduced by including a finite conductivity are of order $v/c \ll
1$; thus, we can self consistently assume infinite conductivity in
Ohm's law.}, ${\bf E}$ is given by the usual Ohm's law.  Substituting
in for ${\bf J}$, we find:
\begin{equation}
\rho \frac{\partial{\bf v}}{\partial t} + \rho {\bf v} \cdot \nabla
{\bf v} = - \nabla P + \frac{1}{4 \pi} (\nabla \times {\bf B}) \times {\bf B} - \frac{1}{4 \pi c} \frac{\partial{\bf E}}{\partial t} \times {\bf B} + \frac{\xi^+}{c} ({\bf v} \times {\bf B}).
\end{equation}
Substituting in for ${\bf E}$ and rearranging leads to:
\begin{equation} 
\rho \frac{\partial{\bf v}}{\partial t} + \rho {\bf v} \cdot \nabla
{\bf v} = - \nabla P - \nabla
\frac{B^2}{8 \pi} + \frac{1}{4 \pi} ({\bf B} \cdot \nabla){\bf B} +
\frac{1}{4 \pi c^2} \frac{\partial{(\bf v} \times {\bf B})}{\partial
t} \times {\bf B} + \frac{\xi^+}{c}({\bf v} \times {\bf B}).
\end{equation}
We use the induction equation to bring ${\bf B}$ into the time
derivative in the penultimate term on the right:
\begin{equation}
\rho \frac{\partial{\bf v}}{\partial t} + \rho {\bf v} \cdot \nabla
{\bf v} = - \nabla P - \nabla
\frac{B^2}{8 \pi} + \frac{1}{4 \pi} ({\bf B} \cdot \nabla){\bf B} +
\frac{1}{4 \pi c^2} \frac{\partial({\bf Q} \times {\bf B})}{\partial
t} - \frac{1}{4 \pi c^2} [\frac{1}{2} \nabla Q^2 - ({\bf Q} \cdot
\nabla){\bf Q}] + \frac{\xi^+}{c} {\bf Q},
\end{equation}
where ${\bf Q} = {\bf v} \times {\bf B}$.  We wish to write the
penultimate term on the right as the divergence of a tensor to
guarantee conservation of angular momentum.  Because $\nabla \cdot
{\bf Q} \ne 0.0$, we must add (and subtract) `$\frac{1}{4 \pi c^2}
{\bf Q} \nabla \cdot {\bf Q}$' to do this. Then, using the equation of
mass conservation to bring $\rho$ into the time derivative on the
left, we find:
\begin{equation}
\frac{\partial \rho {\bf v}}{\partial t} = - \nabla \cdot
\overline{\overline{P}} - \nabla \cdot \overline{\overline{P_D}} +
\frac{1}{4 \pi c^2} \frac{\partial({\bf Q} \times {\bf B})}{\partial
t} - \frac{\xi^+}{c} {\bf Q} - \frac{1}{4 \pi c^2} {\bf Q} \nabla
\cdot {\bf Q} + \frac{\xi^+}{c} {\bf Q} ,
\end{equation}
where 
\begin{equation}
P_{ij}=(P + \frac{B^2}{8 \pi})\delta_{ij} + \rho v_i v_j - \frac{B_i
B_j}{4 \pi},
\end{equation}
and
\begin{equation}
P_{Dij}=\frac{1}{4 \pi c^2} [ Q_i Q_j - \frac{|Q|^2}{2} \delta_{ij} ].
\end{equation}
Note that the extra term we added to the current (the last term in
eq. (A6)) exactly cancels the non-conservation term introduced in
writing $\nabla \cdot \overline{\overline{P_D}}$.  This justifies the
definition of the current in eq. (A2).

We will now neglect the $\overline{\overline{P_D}}$ terms, which are
of order $v/c \ll 1$.  (Note that we can not simply drop the
$\frac{\xi^+}{c} {\bf Q}$ term - even though it is of the same order
of magnitude as the $\overline{\overline{P_D}}$ terms - because it is
needed for conservation.  This is especially important for the current
study because of the key role angular momentum plays in the disk
evolution which is driven by the MRI.)  Canceling these terms, we then
combine the time derivatives and write the mass density as a
generalized tensor:
\begin{equation}
\rho^{\ast}_{ij}=(\rho + \frac{B^2}{4 \pi c^2})\delta_{ij} - \frac{B_i
B_j}{4 \pi c^2},
\end{equation}
to obtain
\begin{equation}
\frac{\partial (\rho^{\ast} \cdot {\bf v})}{\partial t} = - \nabla
\cdot \overline{\overline{P}}.
\end{equation}
The addition of the displacement current enters the momentum equation
only via the mass density.  We apply this change to the ZEUS code by
replacing $\rho$ with $\rho^{\ast}$ for the pressure (gas + magnetic)
source terms.  We then rederive the equations of MOC, following the
method outlined in Hawley \& Stone (\markcite{r}1995; hereafter HS)
for the 1.5 D system.  There is no explicate change in the induction
equation. For the momentum equation, if we ignore the pressure terms
and the off-diagonal terms in the modified mass density, we are left
with
\begin{equation}
\frac{\partial \rho^{\ast} {\bf v}}{\partial t} = - \nabla \cdot (\rho {\bf v} {\bf v}) + \frac{1}{4 \pi} ({\bf B} \cdot \nabla) {\bf B},
\end{equation}
which simplifies to 
\begin{equation}
\frac{\partial \rho^{\ast} (v_x \hat{x} + v_y \hat{y})}{\partial t} = - \frac{\partial \rho v_x (v_x \hat{x} + v_y \hat{y})}{\partial x} + \frac{1}{4 \pi} B_x \frac{\partial B_y}{\partial x} \hat{y}
\end{equation}
for the 1.5 D system. The $\hat{x}$ direction represents only the
advection of momentum.  For the $\hat{y}$ direction, after a little
algebra and making use of the fact that terms of order $v/c \ll 1$, we
can write:
\begin{equation}
\rho^{\ast} \frac{\partial v_y}{\partial t} = - \rho^{\ast} v_x
\frac{\partial v_y}{\partial x} + \frac{1}{4 \pi} B_x \frac{\partial
B_y}{\partial x}.
\end{equation}
Comparing with equation (3.13) of HS, we see that the addition of the
displacement current requires only that we replace $\rho$ with
$\rho^{\ast} = (1+ \frac{B^2}{4 \pi \rho c^2_{lim}}) \rho$ (where
``c'' has been replaced with the limiter ``$c_{lim}$'') in the MOC
equations.  

We note that some of the simulations reported in this paper do not
include the $\xi^+ {\bf v}$ term in eq. [A2].  We find that these runs
show the same initial linear growth and saturation amplitude of the
MRI as the simulations which do include this term.  There is some
indication that the long-term averaged values of the magnetic
quantities (e.g., the magnetic energy) are slightly larger when this
term is included.  However, the large fluctuations which result from
the chaotic nature of the MRI turbulence can lead to even larger
differences in averaged quantities depending, e.g., on the time domain
used in forming the average.  Thus, averages must be performed over
several hundreds of orbital times before secular differences in time
averaged quantities can be considered meaningful.  This suggests that
the difference in the long-term average of the magnetic energy
described above (where ``long-term'' refers to averages over $< 50$
orbits) is not significant.  We stress that the presence of this term
has no effect on the qualitative evolution discussed in the paper nor
on the globally averaged quantities, such as the vertical profiles
(see figs 2, 8, and 11).  There are also no differences in the kinetic
energies, stresses, or fluxes.  The small increase in the time
averaged value of the magnetic energy leads to a 3~$\%$ increase in
the percent of energy dissipated in the coronal region (i.e., the
Poynting flux increases as the magnetic field strength in the disk
increases).  However, we find that a different set of initial (random)
perturbations can lead to differences of order 10~$\%$ in the solution
(again due to the chaotic nature of the turbulence produced by the
MRI) and therefore we do not assign any importance to this 3~$\%$
difference.  In spite of the lack of evidence that the inclusion of
the $\xi^+ {\bf v}$ term in eq. [A2] produces any real change in the
results, we choose to include it in the fiducial toroidal (BY1) and
ZNZ (ZN1) field simulations in order to strictly conserve angular
momentum.  We note that the other runs cited in this paper do not
include this term.

\clearpage

\figcaption[fig1.ps]{Space-time plot showing the vertical distribution
of the horizontally-averaged magnetic (top panel) and kinetic (bottom
panel) energy as a function of time in the fiducial model BY1.  The
magnetic energy varies between 0.0 (black) and 0.846 $P_o$ (red); the
range in kinetic energy between 0.0 (black) and 1.24 $P_o$
(red). \label{fig1}}

\figcaption[fig2.ps]{Vertical variation of horizontally and temporally
averaged quantities in model BY1.  The time average is constructed
between orbits 25 through 50. The horizontal average is taken over the
$\hat{r} \hat{\phi}$ plane as described in the text.\label{fig2}}

\figcaption[fig3.ps]{Space-time plot of $B_xB_y$ (top panel), $\rho
v_x \delta v_y$ ($2^{nd}$ panel), $B_xB_z (3^{rd}$ panel), and
$B_yB_z$ (bottom panel) in model BY1.  The maximum (minimum) value,
colored red (purple), in each plot is 0.322 (-0.118) $P_o$, 0.132
(-0.055) $P_o$, 0.04 (-0.03) $P_o$, and 0.044 (-0.046) $P_o$,
respectively. \label{fig3}}

\figcaption[fig4.ps]{Horizontally and temporally averaged velocity
dispersion in run BY1, normalized by the sound speed. \label{fig4}}

\figcaption[fig5]{Volume averaged $B_y^2$ (solid line), $B_x^2$
(dashed line), $B_z^2$ (dot-dashed line), normalized by $P_o$, within
the disk (left panel) and the corona (right panel) for model
BY1. \label{fig5}}

\figcaption[fig6]{Net magnetic energy generation rate $S_V$ in the
disk (left panel) and the corona (right panel) in model BYR1.  In both
plots, $S_V$ has been volume averaged and normalized by
$P_o$. \label{fig6}}

\figcaption[fig7]{Space-time plot of magnetic energy (top panel),
$B_xB_y$ ($2^{nd}$ panel), kinetic energy ($3^{rd}$ panel), and $\rho
v_x \delta v_y$ (bottom panel) in model ZN1.  The magnetic energy
varies between 0.0 (black) and 0.443 $P_o$ (red), the kinetic between
0.0 (black) and 1.175 (red), $B_xB_y$ between -0.021 $P_o$ (purple) and 0.134 $P_o$ (red), and $\rho v_x$ between -0.030 $P_o$ (purple) and 0.078 $P_o$ (red). \label{fig7}}

\figcaption[fig8]{Vertical variation of horizontally and temporally
averaged quantities in model ZN1.  The time average is constructed
between orbits 10 through 50. The horizontal average is taken over the
$\hat{r} \hat{\phi}$ plane as described in the text.\label{fig8}}

\figcaption[fig9]{Volume averaged $B_y^2$ (solid line), $B_x^2$
(dashed line), $B_z^2$ (dot-dashed line), normalized by $P_o$, within
the disk (left panel) and the corona (right panel) for model
ZN1. \label{fig9}}

\figcaption[fig10]{Space-time plot of magnetic energy (top panel),
$B_xB_y$ ($2^{nd}$ panel), density ($3^{rd}$ panel), and $\rho v_x
\delta v_y$ (bottom panel) in model BZ1.  The magnetic energy varies
between 0.51 $P_o$ (black) and 28.74 $P_o$ (red), the range in density
is 0.0 (black) to 8.69 (red), $B_xB_y$ varies between -1.23 $P_o$
(black) and 20.93 $P_o$ (red), and the range in $\rho v_x \delta v_y$
is -5.1$P_o$ (blacK) to 62.47 $P_o$ (red). \label{fig10]}}

\figcaption[fig11]{Horizontally averaged vertical structure at orbit 5
in model BZ1.  The horizontal average is taken over the $\hat{r}
\hat{\phi}$ plane as described in the text.\label{fig11}}

\figcaption[fig12]{Vertical variation of coronal thermal equilibrium
temperature for model BYR1 for the case of a CTTS. \label{fig12}}

\clearpage

\begin{center}
{\bf TABLE 1:} \\  Parameters of Numerical Simulations  \\
\begin{tabular}{cccccccccc} \hline \hline
\footnotesize
 Run$^{a}$ &
$\beta(0)$ &
$R_M$ &
$c_{lim}$ &
Grid &
$\lambda_c/\Delta$ &
Orbits &
$\langle \frac{B^2}{8 \pi P(0)} \rangle$$^{b}$ &
$\langle \frac{-B_x B_y}{4 \pi P(0)} \rangle$$^{b}$ &
$\langle \frac{\rho v_x \delta v_y}{P(0)} \rangle$$^{b}$ \\
 & & & & & & & ($\times 10^{-2}$) & ($\times 10^{-3}$) & ($\times 10^{-3}$) \\
\hline

BY1 & 25 & $\infty$ & 8 $c_s$ & 32 $\times$ 64 $\times$ 128 & 8.31 & 50 & 8.66/4.16 & 23.6/4.33 & 5.83/0.59 \\ 
BY2$^{c}$ & 25 & $\infty$ & 8 $c_s$ & 32 $\times$ 64 $\times$ 128 & 8.31 & 20 & 6.34/1.80 & 17.8/1.78 & 4.36/0.394 \\
BY3$^{d}$ & 25 & $\infty$ & 8 $c_s$ & 32 $\times$ 64 $\times$ 128 & 8.31 & 20 & 2.61/1.06 & 7.12/1.43 & 2.19/0.220 \\
BY4$^{e}$ & 25 & $\infty$ & 8 $c_s$ & 64 $\times$ 64 $\times$ 128 & 8.31 & 30 & 5.87/3.16 & 15.4/2.57 & 4.37/0.389 \\
BY5 & 25 & $\infty$ & $c_s$ & 32 $\times$ 64 $\times$ 128 & 8.31  & 50 & 5.31/3.10 & 13.7/3.72 & 3.51/0.147 \\
BY6 & 25 & $\infty$ & 4 $c_s$ & 32 $\times$ 64 $\times$ 128 & 8.31 & 15 & 9.41/6.72 & 24.6/9.24 & 6.03/0.629 \\
BYR1 & 25 & 20,000 & 8 $c_s$ & 32 $\times$ 64 $\times$ 128 & 8.31 & 50 & 5.69/3.19 & 14.3/3.17 & 3.87/0.434 \\
BYR2 & 25 & 10,000 & 8 $c_s$ & 32 $\times$ 64 $\times$ 128 & 8.31 & 20 & 5.55/2.98 & 11.1/2.53 & 2.85/0.340 \\
BYR3 & 25 & 20,000 & 8 $c_s$ & 64 $\times$ 128 $\times$ 256 & 16.61 & 20 & 10.9/6.17 & 35.3/3.97 & 8.58/0.582 \\
ZN1 & 25 & $\infty$ & 8 $c_s$ & 32 $\times$ 64 $\times$ 128 & 20.8 & 50 & 2.20/1.28 & 0.49/0.15 & 0.23/0.021 \\
ZN2 & 100 & $\infty$ & $c_s$ & 32 $\times$ 64 $\times$ 128 & 10.38  & 30 & 0.372/0.313 & 0.533/0.600 & 0.269/0.045 \\
ZNR1 & 25 & 20,000 & 8 $c_s$ & 32 $\times$ 64 $\times$ 128 & 20.8 & 15 & 0.486/0.131 & 0.858/0.242 & 1.29/0.046 \\ 
BZ1 & 25 & $\infty$ & 8 $c_s$ & 32 $\times$ 64 $\times$ 128 & 20.8  & 15 & 179.2/101.3 & 907.0/557.2 & 56.9/12.0 \\
BZ2 & 100 & $\infty$ & $c_s$ & 32 $\times$ 64 $\times$ 128 & 10.4  & 10 & 88.1/42.6 & 308.0/115.5 &  35.5/6.99 \\
BZ3 & 100 & $\infty$ & $c_s$ & 32 $\times$ 64 $\times$ 128 & 10.4  & 10 & 103.4/50.1 & 278.2/105.2 & 31.4/2.60 \\
\hline 
\end{tabular}
\end{center}
$^{a}$ Run labels:  the first two letters give the initial field configuration (BY=toroidal field, ZN=zero-net-z field, BZ=pure z field).  ``R'' identifies the resistive runs.\\
$^{b}$ Volume and time averages begun at 15 orbits (BY6, where the average was begun at 10 orbits and ZNR1, where the average was begun at 5 orbits).  The first number represents the average within 2 $H_z$;  the second represents the average for $|z| > 2 H_z$. For the BZ runs, these numbers are volume averages only, computed at orbit 5. \\
$^{c}$ In this run, field was initialized only within $\frac{+}{} 1 H_z$\\
$^{d}$ In this run, $\beta = 25$ at the equator;  B was held constant within $\frac{+}{} 2 H_z$ \\
$^{e}$ In this run, the $\hat{x}$ direction was doubled without increasing the resolution: $ x \epsilon [-1,1]$. \\

\clearpage

\begin{center}
{\bf TABLE 2:} \\  Volume and Time Averaged Quantities in Run BY1  \\
\addvspace{0.5cm}
\begin{tabular}{cccccccccc} \hline \hline
 Quantity f &
$\langle f \rangle$$^{a}$ &
$\langle \delta f^2 \rangle^{1/2}$$^{a}$ &
 min f$^{a}$ &
 max f$^{a}$ \\
 & ($\times 10^{-2}$) & ($\times 10^{-2}$) & ($\times 10^{-2}$) & ($\times 10^{-2}$) \\
\hline

$B_x^2/8 \pi P_o$ & 0.49/0.14 & 0.23/0.09 & 0.0/0.0 & 1.0/0.58 \\
$B_y^2/8 \pi P_o$ & 7.39/3.87 & 2.15/2.58 & 3.92/0.0 & 12.6/8.93 \\
$B_z^2/8 \pi P_o$ & 0.26/0.18 & 0.12/0.12 & 0.0/0.0 & 0.52/0.49 \\ 
$-B_xB_y/4 \pi P_o$ & 2.14/0.43 & 0.97/0.27 & 0.0/0.0 & 3.80/1.47 \\ 
$-B_xB_z/4 \pi P_o$ & 0.01/-0.003 & 0.10/0.04 & -0.09/-0.16 & 0.16/0.20 \\ 
$-B_yB_z/4 \pi P_o$ & -0.03/0.04 & 0.05/0.05 & -0.30/-0.08 & 0.21/0.15 \\ 
$\rho v_x^2/2 P_o$ & 1.11/0.12 & 0.52/0.09 & 0.0/0.0 & 2.65/0.71 \\ 
$\rho \delta v_y^2/2 P_o$ & 0.65/38.8 & 0.28/20.0 & 0.0/0.0 & 1.22/88.3 \\ 
$\rho v_z^2/2  P_o$ & 0.57/0.08 & 2.43/0.06 & 0.0/0.0 & 1.23/0.29 \\ 
$\rho v_x \delta v_y/P_o$ & 0.53/0.06 & 0.28/0.06 & 0.0/-0.17 &  1.36/0.31\\ 
$\rho$ & 64.0/0.71 & 0.37/0.26 & 63.6/0.314 & 64.8/1.28 \\ 
\hline 
\end{tabular}
\end{center}
$^{a}$ The first number represents the average quantity f within $\frac{+}{} 2 H_z$;  the second number represents the average quantity f for $|z| > 2 H_z$. Averages have been taken over 25 orbital times, after saturation.

\clearpage

\begin{center}
{\bf TABLE 3:} \\  Volume and Time Averaged Quantities in Run ZN1  \\
\addvspace{0.5cm}
\begin{tabular}{cccccccccc} \hline \hline
 Quantity f &
$\langle f \rangle $$^{a}$ &
$\langle \delta f^2 \rangle^{1/2}$$^{a}$ &
 min f$^{a}$ &
 max f$^{a}$ \\
 & ($\times 10^{-2}$) & ($\times 10^{-2}$) & ($\times 10^{-2}$) & ($\times 10^{-2}$) \\
\hline

$B_x^2/8 \pi P_o$ & 0.11/0.054 & 0.059/0.036 & 0.0/0.0 & 0.23/0.19 \\
$B_y^2/8 \pi P_o$ & 2.23/1.23 & 1.06/0.67 & 0.0/0.0 &  3.89/1.96 \\
$B_z^2/8 \pi P_o$ & 0.060/0.052 & 0.081/0.036 & 0.01/0.006 &  0.49/0.18 \\
$-B_xB_y/4 \pi P_o$ & 0.54/0.15 & 0.24/0.10 & -0.0004/-0.0001 & 0.95/0.45 \\
$-B_xB_z/4 \pi P_o$ & -0.0034/0.0036 & 0.013/0.017 & -0.059/-0.076 & 0.032/0.065 \\ 
$-B_yB_z/4 \pi P_o$ & 0.0072/-0.0049 & 0.022/0.011 & -0.058/-0.054 & 0.071/0.045 \\
$\rho v_x^2/2 P_o$ & 0.79/0.05 & 0.39/0.039 & 0.0/0.0 & 2.2/0.22 \\ 
$\rho \delta v_y^2/2 P_o$ & 0.27/24.9 & 0.13/12.8 & 0.0/0.0 & 0.66/83.2 \\
$\rho v_z^2/2  P_o$ & 0.21/0.033 & 0.082/0.024 & 0.0/0.0 & 0.40/0.14 \\
$\rho v_x \delta v_y/P_o$ & 0.25/0.022 & 0.21/0.021 & -0.46/-0.07 & 1.26/0.14 \\ 
$\rho$ & 76.8/0.53 & 0.14/0.14 & 76.5/0.29 & 77.1/0.81 \\
\hline 
\end{tabular}
\end{center}
$^{a}$ The first number represents the average quantity f within $\frac{+}{} 2 H_z$;  the second number represents the average quantity f for $|z| > 2 H_z$.  Averages were taken over 25 orbital times, after saturation.

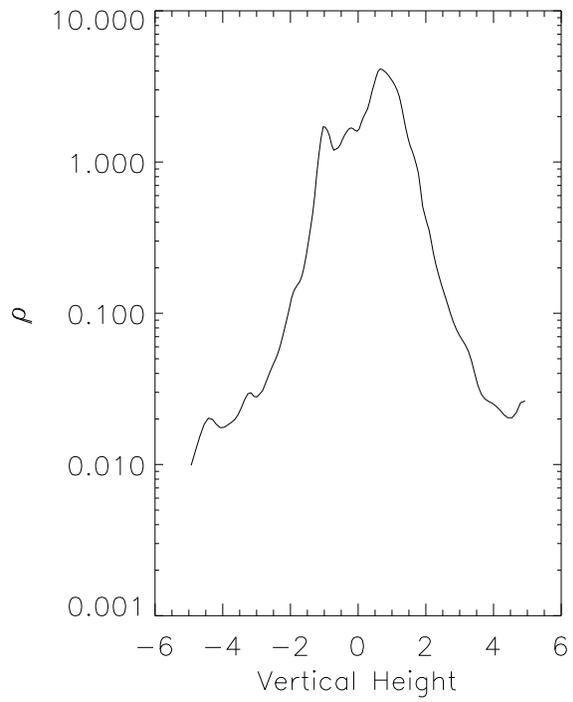
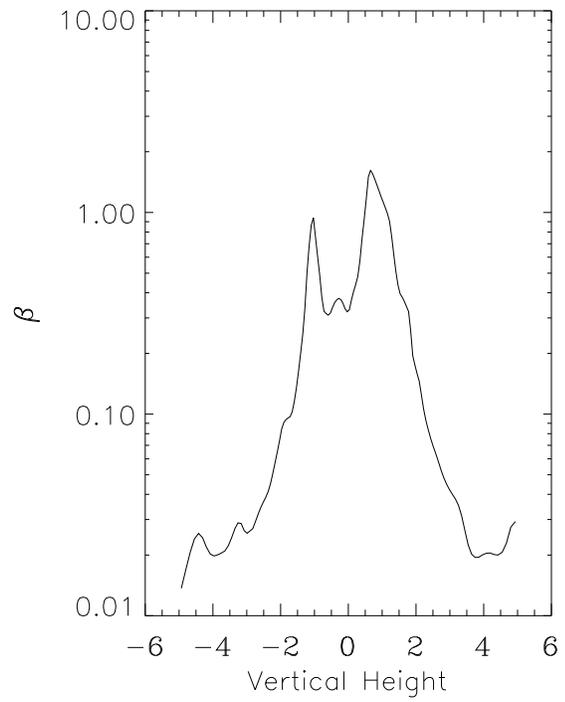
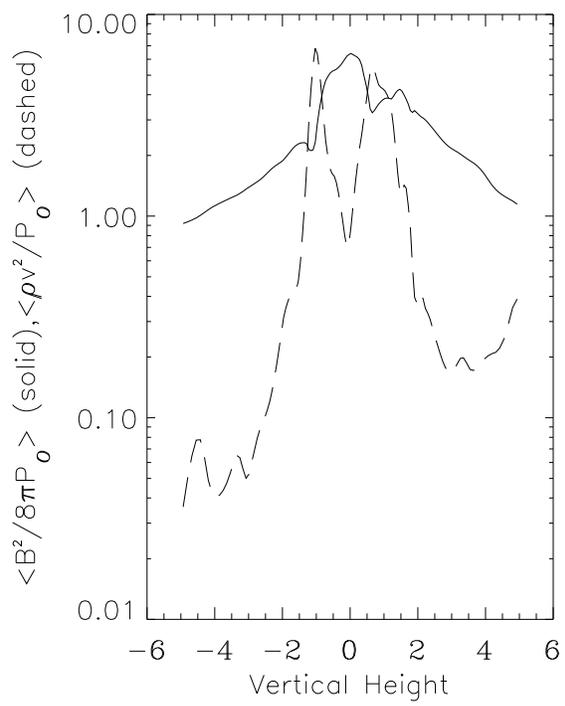
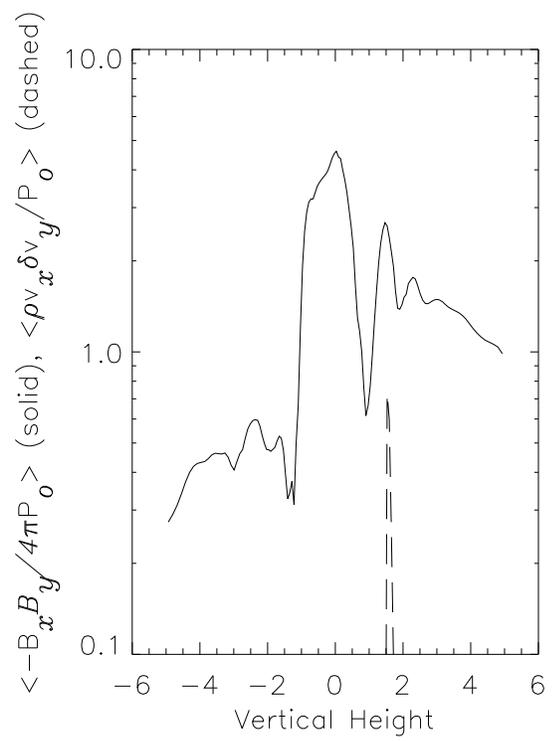
\end{document}